# Coupling Agent-Based Simulations and VR universes: the case of GAMA and Unity

Alexis Drogoul[1,2] [a], Patrick Taillandier[1,2] [b], Arthur Brugière[1,2] [c], Louis Martinez[1], Léon Sillano[1], Baptiste Lesquoy[1,2], Huynh Quang Nghi[2,3]

[1]*ACROSS IJL, IRD / Thuyloi University, 175 Tay Son, Hanoi, Vietnam*
[2]*UMMISCO, IRD / Sorbonne Université, 32 Avenue Henri Varagnat, 93100 Bondy Cedex, France*
[3]*CICT, Can Tho University, 3/2 Street, Ninh Kieu district, Can Tho, Vietnam*
*alexis.drogoul@ird.fr*

Abstract: Agent-based models (ABMs) and video games, including those taking advantage of virtual reality (VR), have undergone a remarkable parallel evolution, achieving impressive levels of complexity and sophistication. This paper argues that while ABMs prioritize scientific analysis and understanding and VR aims for immersive entertainment, they both simulate artificial worlds and can benefit from closer integration. Coupling both approaches indeed opens interesting possibilities for research and development in various fields, and in particular education, at the heart of the SIMPLE project, an EU-funded project on the development of digital tools for awareness raising on environmental issues. However, existing tools often present limitations, including technical complexity, limited functionalities, and lack of interoperability. To address these challenges, we introduce a novel framework for linking GAMA, a popular ABM platform, with Unity, a widely used game engine. This framework enables seamless data exchange, real-time visualization, and user interaction within VR environments, allowing researchers to leverage the strengths of both ABMs and VR for more impactful and engaging simulations. We demonstrate the capabilities of our framework through two prototypes built to highlight its potential in representing and interacting with complex socio-environmental system models. We conclude by emphasizing the importance of continued collaboration between the ABM and VR communities to develop robust, user-friendly tools, paving the way for a new era of collaborative research and immersive experiences in simulations.

## 1 INTRODUCTION

The fields of agent-based modelling (ABM) and video games, including virtual reality (VR), have developed in parallel. Both simulate artificial worlds with rules and agents, but they are used for different purposes. ABMs are used for scientific analysis, while video games aim to provide immersive entertainment. This paper argues that there are fundamental similarities between these two fields and that closer integration can unlock new potential for education and training. This however requires tools that are easy to use and reliable (Berger & Mahdavi 2020).

## 1.1 Convergence

ABMs have emerged as powerful tools for exploring complex systems across diverse disciplines, ranging from social sciences and economics to ecology and urban planning (Drogoul et al. 2002). By modeling individual agents and their interactions within a simulated environment, ABMs offer unparalleled insights into emergent phenomena that arise from the collective behavior of the components of any system. Video games, on the other hand, have matured into intricate virtual worlds capable of captivating users with rich narratives, engaging gameplay, and increasingly immersive experiences (Ivory 2015). VR, the latest frontier in gaming and training, further transcends the screen, transporting users directly into the heart of the simulated world, fostering unprecedented levels of presence and interaction (Bown et al. 2017).

[a] https://orcid.org/0000-0002-9685-9199
[b] https://orcid.org/0000-0003-2939-4827
[c] https://orcid.org/0000-0001-7220-449X

This parallel evolution is not merely coincidental; a closer examination reveals convergences between ABMs and the developments in gaming. Both rely on simulation engines that govern the behavior of agents and the evolution of a simulated world. Both utilize advanced graphics technologies to render realistic environments and more and more engaging visuals. And, especially in advanced ABM platforms like GAMA (Taillandier et al. 2019b), both employ user interfaces that allow users to interact with the virtual space and its virtual inhabitants (Guyot et al. 2006).

Furthermore, the lines between research and entertainment are increasingly blurring (Conesa et al. 2022). ABMs are being employed to inform and enhance game design, leading to more realistic and engaging virtual experiences. Conversely, game engines like Unity (e.g. (Craighead et al. 2008)) or Unreal are being adopted by researchers to build sophisticated immersive research environments for the purpose of serious gaming or participatory simulations (Gazis & Katsiri 2023). This cross-pollination of technologies is further accelerating the convergence between these two fields.

## 1.2 The Potential of ABM-VR Integration

The potential benefits of integrating ABMs with VR platforms are far-reaching and transformative. By leveraging the strengths of both approaches, researchers can:

- Gain deeper insights into complex systems (Juřík et al. 2023), by immersing themselves in the simulated world alongside the agents, observing their interactions and behaviors firsthand. This allows for a more intuitive and visceral understanding of the system's dynamics.
- Conduct more realistic and controlled experiments (Lin et al. 1999): Design virtual experiments that closely mimic real-world scenarios, allowing for the exploration of different hypotheses and the testing of interventions in a risk-free environment.
- Enhance engagement and understanding for diverse audiences: Communicate research findings and complex concepts through immersive and interactive experiences, thereby making them more accessible and engaging for a broader audience.
- Revolutionize education and training (Richards 2008; Brasil 2011): Develop virtual learning environments that offer interactive and personalized learning experiences, fostering deeper engagement and understanding of complex subjects.

These applications are far from theoretical. Researchers are already utilizing ABM-VR integration in a variety of domains, including:

- Social science: Studying collective behavior, crowd dynamics (Shendarkar et al. 2008; Liu et al. 2014).
- Urban planning: Experiencing the impact of different urban design decisions on the built environment and its inhabitants (Zhang et al. 2009; Yu et al. 2014).
- Ecology: Exploring the interactions between different species and the impact of human intervention on ecosystems (Cho & Park 2023).
- Healthcare: Training medical professionals in a safe and controlled virtual environment (Possik et al. 2021; Possik et al. 2022).
- Education: Learning about history, science, and other complex subjects through interactive and immersive experiences (Lin et al. 1999; Popovici et al. 2004; Bosse et al. 2014).

## 1.3 Education to Sustainable Development: the SIMPLE Project

Education, the final point, is at the heart of the SIMPLE project[4], funded by the European Union as part of the EU-ASEAN Green Partnership, which aims to explore the relevance of educational virtual environments for raising teenagers' awareness of sustainable development and environmental issues. It is based on the idea that integrating ABM-VR into the curricula can transform the learning experience on these large-scale, complex issues, which are not generally amenable to experiential learning.

In the project, on one hand, ABMs are used for integrating the knowledge and concepts from various disciplines, something required for representing the evolution of socio-environmental issues as complex as the subsidence of coastal areas, the loss of biodiversity of forests, the leakage of plastics in the ocean or the importance of agro-ecological practices for sustainable agriculture. On the other, interactive VR simulations, based on scientific models, can offer immersive and engaging ways for young people to explore complex topics, fostering deeper understanding through virtual experiences and encouraging a new generation of critical thinkers and problem solvers (Sancar et al. 2023).

---

[4] https://project-simple.eu

SIMPLE will support the design, implementation, and experimentation in classrooms of at least six ABM-backed virtual universes in Vietnam, Thailand, Lao and Cambodia from 2024 to 2026, on topics co-constructed by the partners of the project, local research institutions and educational communities.

## 1.4 The challenges of a seamless integration between ABM and VR

Despite the undeniable potential, integrating ABM and VR platforms still presents significant challenges. Existing approaches (Louloudi & Klügl 2012; Huang et al. 2018; Adinolf et al. 2019) often suffer from technical complexity, limited functionalities, and lack of interoperability between ABM platforms and game engines. Furthermore, the literature on the subject is quite scarce. For instance, (Ospina-Bhorquez et al. 2021) provides a systematic mapping of the literature exploring synergies between multi-agent systems (MAS) and virtual reality environments. Although MAS is a broader field of application than ABM, it is interesting to note that (1) they only found 82 articles of interest; (2) the vast majority of papers focus on applications rather than frameworks or platforms (indeed, this review is mainly organized in terms of application domains, most of which, interestingly, relate to simulation domains such as urban traffic, crowd simulation, or robot control).

For example, in (Louloudi & Klügl 2012), the general framework for coupling an agent-based platform (SeSaM) with a 3-dimensional rendering tool (Horde3D) ultimately addresses only the issue of visualizing simulations and not that of mapping interactions between agents and player(s) in the two environments. In (Huang et al 2018), which addresses, in the context of urban design, the issue of coupling not only visualization but also behavior, agent-based paradigms are ported within Unity, with the authors using the C# scripting engine to write an agent-based model. This is also the path chosen by other authors, in particular (Conesa et al. 2022), also in Unity, by considering the GameObjects offered by Unity as autonomous agents, (Olivier et al. 2014), dedicated to crowd simulation, or (Kamalasanan et al. 2022), dedicated to cycling in realistic traffics. In contrast to these approaches, which are more akin to porting than coupling, (Possik et al. 2022) represents one of the most successful attempts at operational coupling, based on three pieces of software: Unity (VR) and AnyLogic (ABM), as well as Papyrus, the latter serving to orchestrate the whole. All three are linked using the HLA platform, originally designed to facilitate the design of distributed, modular simulations. However, this work, which is based on truly interesting elements, has not necessarily been designed, for the time being, to be reusable. The software complexity of the whole package is furthermore considerable and making it available or reusing it in contexts other than the propagation of Covid19 in a hospital, for which it was designed, does not seem to be one of the authors' objectives.

In this paper, we introduce the SIMPLE platform, a framework for easily linking GAMA, a leading ABM platform, with Unity, a widely used game engine, which addresses the limitations of existing approaches by:

- Providing a software solution that's independent of the application domain: neither GAMA nor Unity are tied to specific domains, so the SIMPLE platform isn't either, instead relying on the use of operational abstractions.
- Providing a user-friendly interface and a modeler-friendly access to VR: the SIMPLE platform simplifies the process of coupling GAMA and Unity, allowing researchers with diverse backgrounds to leverage the framework.
- Enabling seamless and flexible data exchange: it facilitates the real-time transfer of data between the ABM and VR environments, ensuring accurate and consistent visualization of the simulated system, and offers tools to monitor these exchanges.
- Supporting user interaction: it allows to describe VR input into ABMs, allowing players in the virtual universes to interact with the simulated worlds and influence the behavior of agents in the ABM in real-time.

This framework represents a significant step forward in bridging the gap between ABMs and VR. It empowers researchers to create immersive and interactive virtual environments that enhance understanding, facilitate experimentation, and engage diverse audiences.

## 2 FLEXIBLE COUPLING BETWEEN GAMA AND UNITY

GAMA has always been at the forefront of agent-based modeling platforms in terms of visualization (Grignard and Drogoul, 2017) and interaction between users and simulations. Offering early on the possibility of multiplying viewpoints on a model, either via multiple 2D and 3D displays, or via specific devices (Brugière et al. 2019), the platform has earned itself a fine reputation in the field of participatory simulations (Taillandier et al. 2019a).

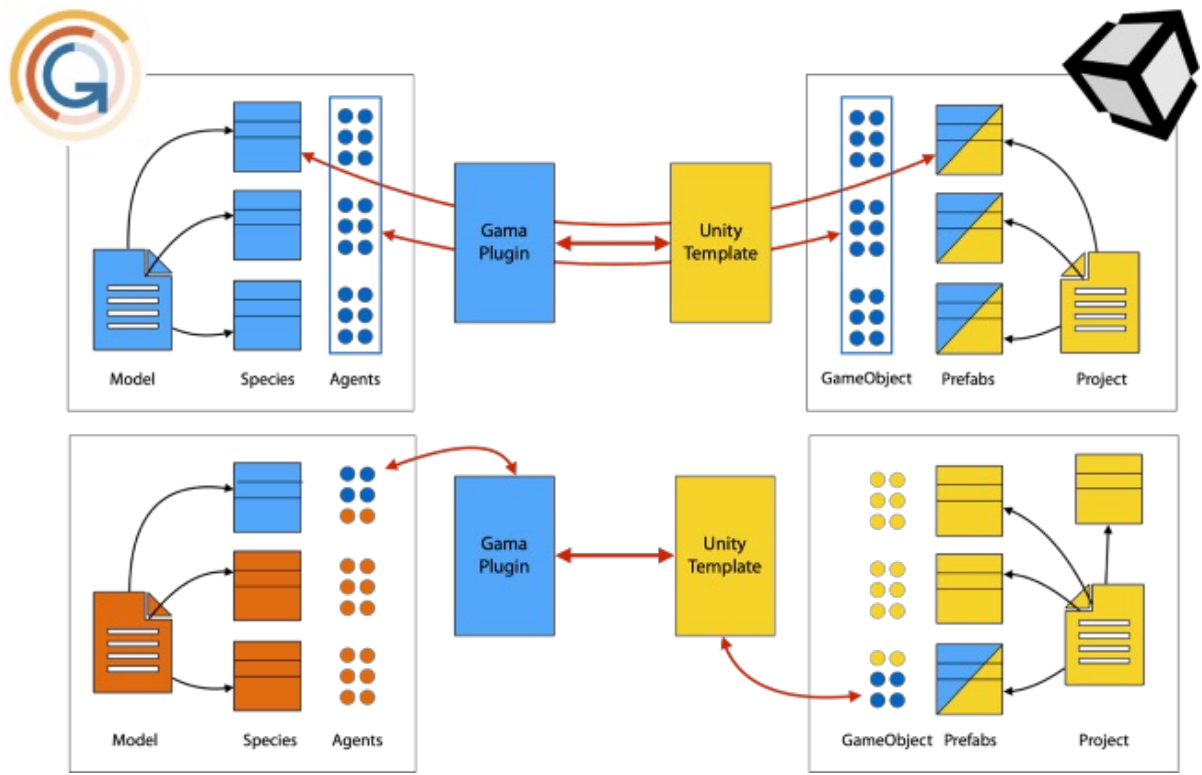

Figure 1 Coupling is based on the definition of a plugin (on the GAMA side) and a model (on the Unity side). This combination allows a wide range of configurations, from bijection (top) to incomplete projection (bottom), to the simple exchange of data between the platforms.

At the heart of GAMA lies the concept of "agent", an instance of a "species," which acts analogously to classes in object-oriented programming, serving as a blueprint for individual agents within the simulation, defining their behaviors, attributes, and interactions. Similarly, Unity's fundamental building block is the "GameObject," a versatile entity representing any visual or interactive element within the game engine, which can be instantiated from a "prefab". Prefabs are pre-configured sets of GameObject components that can be easily replicated and customized, allowing developers to create large numbers of visually consistent agents with minimal effort.

Built on this conceptual similarity, the coupling between GAMA and Unity aims to seamlessly translate the GAMA concepts into Unity's environment, enabling the creation of immersive and interactive virtual worlds powered by agent-based modelling. However, the translation process requires careful consideration of the diverse nature of agents and the desired level of integration between the model(s) and the virtual universes built on top of them. We have identified three potential modes of coupling (Fig. 1), of course not exhaustive:

- **Bijection**: In this mode, each GAMA agent is directly translated into a corresponding Unity GameObject. The GAMA species acts as a template for a prefab, which in turns defines GameObjects' behaviors and attributes (but not their visual appearance). This approach offers the highest level of fidelity, allowing for complete control, in GAMA, over the individual representation of agents within the virtual world, using Unity as an interface of the simulation.
- **Projection**: A more flexible approach, required by some models. Instead of translating every GAMA agent into a GameObject, only a subset deemed essential for visual representation or interaction is converted. Other agents may not be explicitly represented visually but remain active in the simulation, contributing to background dynamics and influencing other agent behaviors.

This approach optimizes performance while still maintaining the core functioning of the model.
- **Background**: In this mode, GAMA acts as a background simulation engine, driving the underlying dynamics and behaviors of the virtual universe without having any direct visual representation in Unity. This is suitable for situations where the virtual universe relies on the model for data and computations, and where the influence of the agents on the environment or other agents needs to be accounted for.

As can be seen above, SIMPLE platform's coupling between GAMA and Unity extends beyond simply translating agents into game objects. The framework also enables the exchange of data and information between the two platforms, allowing for real-time updates and dynamic interactions within the virtual environment and a great flexibility in how the two worlds communicate.

## 2.1 On the ABM side: a dedicated GAMA plugin

We have developed a plugin for GAMA to provide a connection with Unity or Unity-based products. This plugin extends a GAMA simulation by enabling the connection of external clients, like players using VR headsets. This connection can be used in two modes of operation: direct or indirect. The direct mode establishes a direct connection between a VR headset and GAMA but does not allow more than one player to be connected. The indirect mode requires the use of another software (the GAMA Server Middleware, available and documented at https://github.com/project-SIMPLE), but enables multiple players to be connected simultaneously. The middleware also provides additional monitoring and management tools of the connections. Both methods use the same GAMA Server API, which not only enables messages to be sent in both directions, but also allows clients to control a simulation by directly executing code defined in its model. Configuring it is easy, with just a few highly configurable species of agents to add to establish the connection.

The first species of agents introduced by the plugin, *abstract_unity_player*, offers an optional representation of the players in the model. It allows to track and visualize them (if their representation in the model makes any sense) but also to provide them with behaviours or capacities of interaction with other agents in the simulation. A second species, a key element of the plugin, is *abstract_unity_linker*. It allows the creation of agents that connect GAMA and a game running on Unity. More precisely, it oversees creating player agents when a player connects, sending the elements needed to initialize the virtual world (world size, geometries, etc.) and, at each simulation step, synchronizing the information defined by the modeler between the two worlds (for instance, the location and orientation of agents, etc.). The last species allows to instantiate a new type of experiment (*unity*) that automatically creates and initializes a corresponding *abstract_unity_linker* agent based on the parameters entered by the modeler.

These species form the basis of the components injected by the plugin into GAMA. They can be used directly, like any agent, when creating models to support VR games. However, as one of the most likely use cases for SIMPLE is to connect existing models to VR games, we have designed a wizard for this purpose. This allows modelers to enter various parameters and generates a derived model in which all these components are correctly initialized, enabling for instance the exploration of simulations of the source model using a VR headset almost instantly and without any special knowledge of Unity.

## 2.2 On the VR side: a dedicated Unity template

Another significant development in the SIMPLE platform is the Unity template that offers *prefab managers* to handle the connection with GAMA, as well as pre-configured Unity projects with plugins providing different properties (physics-aware movements, etc.). It also provides three scenes to facilitate game construction and support the interaction using the controllers of the VR headset:

o *Start Scene*: a simple main menu that allows to load two other scenes - the *IP Menu Scene* and the *Main Scene*. This menu also allows defining whether the middleware will be used to connect to GAMA or not.
o *IP Scene*: allows the user to change the address used to connect to the computer running the middleware or GAMA.
o *Main Scene*: main scene composed by default of the following *GameObjects*:
  o *Directional light*: default light for Unity scenes. This type of light corresponds to a large, distant source originating from outside the range of the world.
  o *Player*: predefined GameObject with First Person View and movement for VR. This package offers different types of player

settings. For instance, in games requiring large-scale decision-making, the package provides a type of player who can fly over the game space
- *Connection Manager:* main interface with GAMA, sets Unity connection settings to GAMA in this package.
- *Game Manager*: main coupling interface with GAMA. It defines all aspects of the game, such as how to represent the agents in the game or how to convert space coordinates from GAMA to Unity.
- *Debug Overlay* (can be disabled): used to show messages from GAMA in the headsets for debugging purposes.

The software package has been designed to provide a high degree of configurability and ease in replacement of components, with the aim of accommodating a diverse array of use cases. Beyond accommodating various player types, the package allows for the explicit definition of player interactions with the simulation from the game. This is achieved through two main modes of interaction: triggering actions through an agent (which can be any agent within the simulation, not exclusively the player agent), or sending GAML code directly executed by GAMA. The package also facilitates processing messages received from GAMA in JSON format. It manages the deserialization of messages and subsequently initiates actions as per the communicated instructions. Messages can range from simple function calls to more intricate transmissions, such as the dissemination of the positional information of all agents within the simulation. Moreover, this package encompasses two editor-level tools, both based on the connection with GAMA. The first one facilitates the importation of geometries from GAMA in either two or three dimensions and their translation to *GameObjects*. This feature allows to retrieve geographic data, initially read by GAMA (such as shapefiles or other geographical data), while maintaining the same coordinate system. Then, users can manipulate the visual representation of these geometries within the Unity environment. The second one operates in the reverse direction, exporting *GameObjects* from Unity to GAMA and saving them as shapefiles (in 2D). This feature facilitates the connection between virtual worlds developed in Unity and GAMA simulations.

## 2.3 Using ABM and VR extensions together.

The tools we developed were designed to enable modelers to quickly build a first prototype VR version of an existing GAMA model with minimal knowledge about Unity and VR technologies. The tools have been developed for GAMA 1.9.3 (https://gama-platform.org/download) and Unity 2022.3.5f1 (https://unity.com/download). They require GAMA to install the Unity plugin and Unity to use the dedicated template project, both of which are available on SIMPLE's GitHub repository (https://github.com/project-SIMPLE).

For the sake of this section, we'll assume that the modeler has already developed a GAMA model and wishes to implement a VR headset interaction with this model. Three steps are needed to create a Unity environment from an existing GAMA model:

1. Generation of a GAMA model extending the base model with VR capabilities.
2. Development of the virtual universe in Unity.
3. Implementation of specific actions that VR players can execute in the GAMA simulation.

### 2.3.1 Step 1. Generation of the GAMA model for connection to Unity.

Upon installation, the Unity template for GAMA automatically generates a new GAMA model that extends the base model and allows its connection to a VR headset. To set this up, the user needs to make a few choices. The first concerns the number of expected players in the simulation. It is thus possible to define a minimum number of players to launch the simulation and a maximum number who will be able to connect to it. The user also needs to choose the initial experiment for the extended model, pick how the player will be shown, and specify any displays not to be used from the initial experiment. The next step is to select the agent species whose coordinates will be sent to Unity at each time step. Another choice to make is to define the agent species whose geometries will be sent to Unity at initialization. The aim here is to send geometries that will not be modified during the simulation, such as the buildings or roads in a traffic model. For each of these species, it is possible to define whether they will be in Unity 2D or 3D, their colour, an associated tag, whether they will have a collider and whether it will be possible to interact with them.

### 2.3.2 Step 2. Development of a virtual universe under Unity.

Once the basic model is generated in GAMA, the next step consists in developing the virtual universe. The modeler can do this by using the three scenes available in the SIMPLE template and parametrizing the main scene accordingly. If the modeler keeps the default First Person View-type player, the sole parameterization required concerns the Game manager. Specifically, for each species dispatched from GAMA, the modeler must specify their representation within the Unity environment. This involves determining the appropriate prefab, establishing the scaling factor, deciding on the application of rotation or translation, and addressing related considerations. The Game manager thus serves as the central locus for configuring the visual representation and spatial attributes of each agent species within the Unity interface.

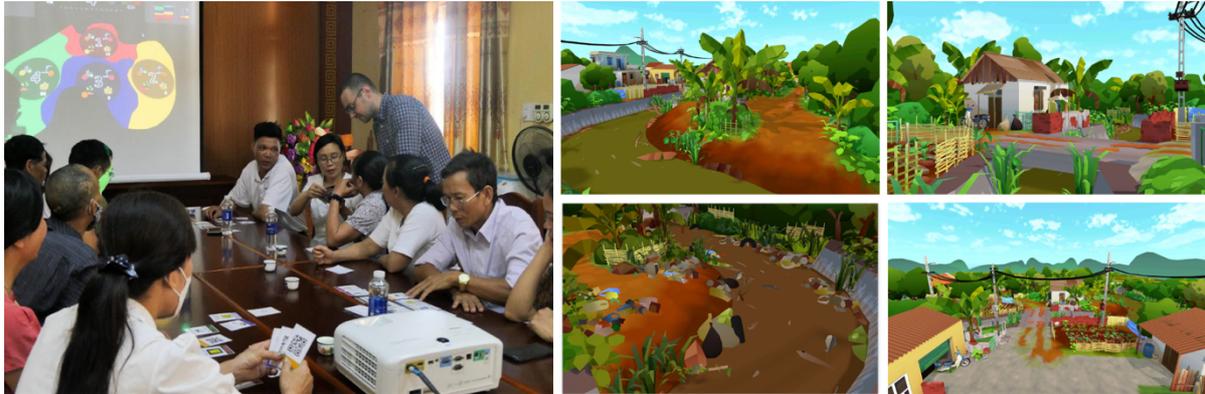

Figure 2 - Left shows a game session of the card version of the Rác model, right shows some in-game images of the VR adaptation of the model.

### 2.3.3 Step 3. Definition of specific interactions

The last, optional, step consists of defining specific interactions between Unity and GAMA. The default SIMPLE template for Unity already provides illustrative instances of interactions, such as modifying the lighting through button presses on controllers and enabling GAMA actions triggered when a player selects an object with a ray interactor.

To make it easier, it is possible to use these existing examples and adapt them to newer interaction with the extended model.

For instance, in a traffic model, it is possible to define in GAMA (within the plugin's *Unity Linker* library) an action to close a road, with the road name as an argument. It will then just be necessary in the script linked to the *Game Manager* in Unity to specify the name of this action when a road is selected with the name of the road's *GameObject* as an argument. It is of course possible to go further in the parameterization of actions by adding actions linked to default Unity's interfaces (as the OpenXR's *XRI Default Input Actions*) which will also trigger specific actions in GAMA from controller's interaction.

## 3 VALIDATIONS

While the two first virtual universes planned in the SIMPLE project are still in pre-production, we have begun to validate crucial aspects of the platform with the design of two prototypes that extend existing agent-based models and explore different interactions with users in virtual worlds linked to simulations.

### 3.1 RÁC VR: a VR extension to a serious game based on an ABM.

RÁC is a multiplayer game designed to raise awareness of the issue of domestic waste pollution and its impact on an irrigation system in Vietnam. It also aims to foster social dialogue between the various stakeholders involved in this pollution.

The game follows the evolution of waste management and agricultural production in the fictitious territory of a rural commune. This commune is divided into four villages, each facing specific problems due to its upstream or downstream location.

The game is played by four groups of two or three players (8 to 12 players in total, see Fig. 2). Each group represents a village chief and makes decisions on actions that will have a local impact to achieve a common goal for all four villages: maintaining the so-called *EcoLabel* certification. This fictitious label, which refers directly to the *VietGap* quality label, is presented as necessary for farmers to be able to sell their produce to supermarkets. To retain *EcoLabel* certification, the commune must achieve a minimum level of agricultural production and a limited level of soil and water pollution on its fictitious territory, meaning that each village must meet these requirements individually so that *EcoLabel* certification can be guaranteed, collectively, for the commune. All players are then invited not only to improve the situation in their village, but also to coordinate their actions to influence and promote agricultural productivity at commune level. The game is based on two materials:

- Cards: each group receives a complete deck of cards. Each card represents an action linked to waste management or agricultural production, with a QR code.
- A GAMA computer simulation: this simulates the evolution of the territory, allowing players to observe the consequences of their actions.

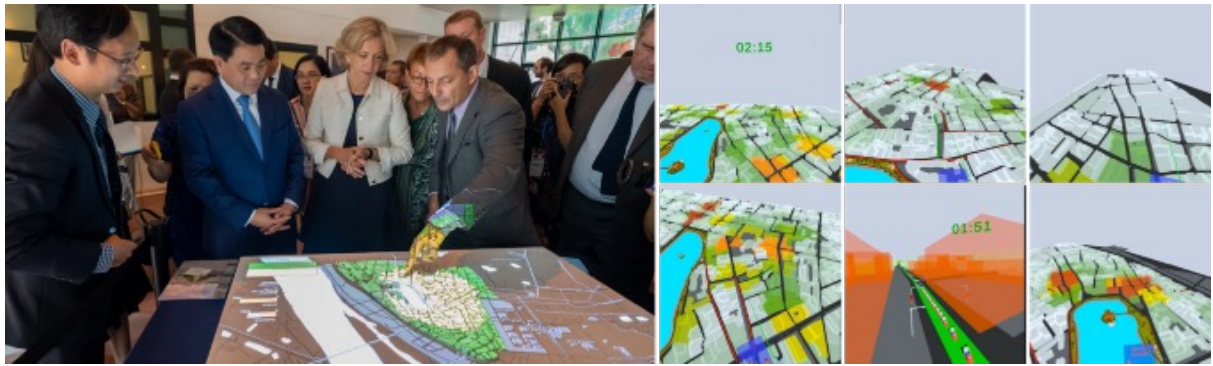

Figure 3 - Left: presentation of Hoan Kiem Air at the French Embassy in Hanoi; right: some in-game images of the VR adaptation. The colors of buildings show the level of pollution. Roads are displayed in red when closed, black when opened and green when selected.

RÁC was played in two contexts: with high-school students and with farmers and village leaders. The first series of games with students, which mainly served to test the game, showed the importance of the emergence of leaders among the players to ensure the coordination of decisions taken and therefore the effectiveness of the chosen policy. The second series of experiments with farmers and village chiefs, in addition to taking part in the debate on waste management in their territory, highlighted discussion regimes and depicted specific circulations of power within different categories of actors.

The VR version (Fig. 2 right) is based on the same principles as RÁC, but with an emphasis on raising awareness of pollution issues among young people. This version uses four Meta Quest 3 headsets directly connected to GAMA. The idea was to add an immersion phase in the villages to make players more aware of the impact of pollution, and to force them to discuss their own perceptions of this pollution. More specifically, before each decision-making phase, during which groups of players choose which actions to take, a village exploration phase is added. During this phase, one of the players in each group will put on the VR headset to explore "their" fictional village. If this village is not directly linked to the territory simulated in GAMA, its characteristics are consistent with it. GAMA will send information on pollution and agricultural production to each VR headset at the start of the VR phase, representing the situation of each player's village. These levels of pollution (solid and water) and production will be represented in the VR headset (solid waste, color of water, color of plants) allowing the player to observe directly by immersion the impact of decisions taken at village level. The challenge for the immersed player will then be to describe to the other players what he or she has seen, so that they can then make the right decisions: typically, if the player observes and describes a high level of waste, the other players may take actions linked with solid waste management. Conversely, a map representing the village to explore is displayed on the GAMA screen, along with the players' positions, which allows players without VR headsets to give advice on the direction to follow to reach interesting points to explore (e.g. fields, river, etc.).

In this prototype, we are therefore in the case of "Background coupling", where the simulated environment (the commune) is different from the one seen by the players in the VR headset. Only information on pollution and production levels is sent by GAMA, and Unity only sends the player's position back to GAMA so that other players can follow what's happening for the VR headset-equipped player.

In the context of this prototype where the link between GAMA and the VR headsets is very simple and where only some information passes, the only thing to define was the information sent by GAMA at a given moment in the simulation. This information includes three values: solid pollution, water pollution and agricultural production. From the moment the headset receives this information, the exploration phase of the village in VR can begin. A timer is present in the Unity application to limit exploration time. Once this timer ends, the headset sends a message to GAMA to specify that the VR exploration phase is over for the player. In this prototype, there is therefore no species of agents sent at each simulation step to the headset or geometries used to define the environment. There is also no interaction in the headset that impacts the simulation in GAMA.

### 3.2 HoanKiem VR: a transposition of a tangible ABM interface into VR

The Hoan Kiem Air model was designed to open discussions on the introduction of pedestrian zones in terms of traffic and air pollution. Many cities have seen the emergence of pedestrian zones in recent years. If the interest of such zones is multiple, they can also have negative impacts due to the deferral of traffic in terms of congestion and therefore air pollution. The Hoan Kiem Air model thus makes it possible to simulate traffic and pollution emitted by vehicles in the Hoan Kiem district of Hanoi, Vietnam. In this district, the surroundings of the central lake are closed to traffic every weekend and the local authorities have been considering extending the pedestrian zone for several years. The Hoan Kiem Air model allows to simulate these different scenarios (without pedestrian zone, with current pedestrian zone or with extended pedestrian zone) and to see their impact on traffic and pollution. The heart of the model is based on the traffic plugin included in GAMA (Saval et al. 2023), which enables to simulate road traffic made up of different types of vehicles (cars, motorcycles, etc.) to be finely represented at the scale of a district or city.

To facilitate discussions around the simulations, Hoan Kiem Air also includes a 3D physical model of the district onto which the simulation is projected. The use of a tangible interface makes it possible to better grasp the spaces, give more reality to the outcomes and facilitate exchanges by placing the stakeholders all around the physical model (Fig. 3 left). Finally, an Android application (running on a tablet) directly connected to GAMA allows users to control the parameters (scenario for pedestrian zones, number of vehicles, etc.) and by this to quickly test different configurations.

The VR version of Hoan Kiem Air (Fig. 3 right) allows us to go further in the possibilities of interaction with the simulation by also providing a playful aspect intended for a younger audience. It thus allows a player equipped with a VR headset to close/open roads themselves and directly see the impact (both positive or negative) of their choices on traffic and pollution, materialized by buildings' color and a score. The challenge is therefore for the player to have the best possible scores by judiciously choosing the roads to close. In practice, the game has the main advantage of showing that it can be complex to limit pollution only by closing roads due to the deferral of traffic.

In the VR headset, the player can observe the entire district from above: the player can see all the buildings, roads and vehicles already present in the GAMA simulation. Likewise, in the GAMA simulation, other people can observe the player's position, his orientation (where he is looking) as well as the roads he has decided to close. In this prototype, we are therefore in the case of "Bijective coupling", where all the information in the GAMA simulation is displayed in the headset.

To be more precise, 2 species are sent from GAMA to the VR headset at each simulation step (i.e. they are part of the list of species to be sent to Unity from the simulation defined in the GAMA model): cars and motorcycles. We chose 2 prefabs in Unity for each of these species, representing a typical car and a typical motorcycle in Hanoi. To accommodate the limited capabilities of the headsets and ensure the display of a substantial number of agents while maintaining a smooth framerate, low-polygon assets were employed for rendering these two types of objects. In addition, roads and buildings rendered in the VR headset are sent by the GAMA model at initialization. As the aim was to enable roads to be closed, we made them interactable by specifying, in the GAMA model, that they have a collider in VR. Finally, we specified in Unity the GAMA action to be called when a specific road is closed (defined in the *Unity Linker* of the GAMA plugin).

## 4 CONCLUSION

In this article, we introduced the SIMPLE platform, which bridges the gap between GAMA, a popular agent-based modelling platform, and Unity, the well-known game engine. We first gave an overview of the relationships between the concepts used in GAMA and Unity, and then presented how we link these concepts between the two platforms and how their coupling can be achieved via SIMPLE. Examples from two prototypes (RAC and HoanKiem Air) are presented and discussed.

### 4.1 Limitations

While SIMPLE offers a valuable bridge between GAMA and Unity, limitations remain that hinder the full potential of their integration. These limitations primarily revolve around two key areas: a reliance on manual steps and the lack of a dedicated process to simplify the workflow for modelers.

The current manual configuration for coupling GAMA species with Unity *GameObjects* can be tedious and prone to errors. Manually mapping attributes, defining behaviors, and setting up interaction mechanics is time-consuming and can discourage adoption, particularly for users less familiar with both platforms. Automating as many of these steps as possible, potentially through code generation or drag-and-drop interfaces, would significantly improve accessibility and efficiency.

The lack of wizards is a particularly critical missing element. Providing guided workflows tailored to specific use cases would greatly benefit modelers. These wizards could handle common configurations, suggest relevant Unity components, and automate repetitive tasks, allowing modelers to focus on the core logic and design of their simulations.

## 4.2 Developments

Addressing these limitations through enhanced automation and user-friendly tools is crucial. This will involve:

- Providing a GAMA runtime endpoint directly in the Unity extension: Allowing Unity to make requests to GAMA and directly compile experiments able to be linked to Unity.
- Developing intelligent mapping tools: Automatically identifying corresponding attributes and behaviors between GAMA species and Unity GameObjects, minimizing manual configuration.
- Implementing context-aware wizards: Providing guidance based on the model being developed, offering relevant options and best practices.
- Exploring plug-and-play components: Building libraries of pre-configured Unity components designed to work seamlessly with specific GAMA species functionalities.

By overcoming these limitations, SIMPLE will evolve into a truly streamlined and accessible tool for coupling GAMA and Unity. This will not only simplify the modelers' workflow, but also encourage wider adoption and unlock the potential of this integration for research, education, and virtual world creation. An important point will be to support each major release of both development environments, which will be achieved through the first point in the list above (i.e. by integrating a version of GAMA into each version of the Unity extensions).

## 4.3 Future developments

While addressing the limitations of SIMPLE represents a critical step, the future of GAMA-Unity coupling holds even more exciting possibilities. Here, generative AI (Becker et al. 2023; Qin & Hui 2023; Chamola et al. 2023) emerges as a promising tool to ease the way we bridge these two platforms and we have already begun to explore it in a series of research works.

Generative AI can significantly reduce the reliance on manual configurations, currently a major bottleneck in the coupling process. Imagine intelligent algorithms automatically mapping GAMA species attributes to Unity components, suggesting optimal configurations for behaviors and interactions, and even generating code snippets for custom functionalities. This level of automation would streamline the workflow, allowing modelers to focus on the bigger picture without getting bogged down in technical details.

Generative AI can also unlock dynamic content creation within virtual environments (Ratican et al. 2023). Imagine a system that automatically generates diverse and unique Unity assets, such as buildings, landscapes, and even characters, based on the underlying GAMA simulation parameters or natural language prompts used for the aspects of agents. This would support fast prototyping of complete virtual worlds, offering an unparalleled way to test ideas and offering modelers without a graphics expertise the possibility to deploy appealing virtual worlds within minutes.

By embracing the power of generative AI on top of the existing coupling mechanisms, with intelligent automation and dynamic content creation, we aim at transforming the SIMPLE platform into a truly transformative tool for research, education, and -- why not -- entertainment.

## ACKNOWLEDGEMENTS

This publication was produced with the financial support of the European Union. Its contents are the sole responsibility of the authors and do not necessarily reflect the views of the European Union.